\documentclass[letterpaper,aps,prl,superscriptaddress,showpacs,floatfix,twocolumn]{revtex4}

\usepackage{graphicx}
\usepackage{hyperref}
\usepackage{amsmath}
\usepackage{url}
\usepackage{colordvi}
\usepackage{subfigure}

\usepackage{times}

\begin{document}

\title{Valence Quark and Gluon Distributions of Kaon from $J/\psi$ Production}

\author{Jen-Chieh Peng}
\affiliation{Department of Physics, University of Illinois at
Urbana-Champaign, Urbana, Illinois 61801, USA}

\author{Wen-Chen Chang}
\affiliation{Institute of Physics, Academia Sinica, Taipei 11529, Taiwan}

\author{Stephane Platchkov}
\affiliation{IRFU, CEA, Universit\`{e} Paris-Saclay, 91191 Gif-sur-Yvette,
France}

\author{Takahiro Sawada}
\affiliation{Department of Physics, University of Michigan,
Ann Arbor, Michigan 48109-5586, USA}
\date{\today}
\begin{abstract}
The only experimental information on the parton distribution of
kaons was obtained from the scarce kaon-induced Drell-Yan cross
section data. From these data, evidence for a valence $u$ quark 
distribution of kaon
softer than that of pion was found. We study the feasibility to
extract kaon's valence quark distribution via the 
existing kaon-induced $J/\psi$ production data. We compare the
NA3 data on the $K^- / \pi^-$ and 
$K^+ / \pi^+$ ratios for $J/\psi$ production on a platinum target with 
theoretical calculations based on the color-evaporation
model. We show that the NA3 data on $J/\psi$ production provide
an independent evidence for a valence $u$ quark distribution softer in
kaon than in pion. The data also suggest different gluon distributions
for kaon and pion. 
\end{abstract}

\maketitle

Since the discovery of partonic structures of hadrons in deep inelastic
scattering (DIS), extensive theoretical and experimental progress has been
made in our knowledge on proton's partonic distributions. In contrast,
the partonic structures of pion and kaon, which have the
dual roles of the lightest quark-antiquark bound states and the Goldstone
bosons, remain poorly known. Significant theoretical efforts 
have been devoted to predicting the quark and gluon distributions 
in pion and kaon, using QCD-based frameworks~\cite{Roberts_Holt}. 
These predictions remain 
to be tested against new experimental information.

As mesons are not available as targets for performing DIS experiments,
the existing experimental inputs for the parton structures of
mesons almost exclusively come from Drell-Yan process~\cite{DY,Kenyon,
WA39,NA3,Chang13} and direct photon
production~\cite{WA70} with meson beams. These data led to the 
extraction of pion's valence
quark distribution~\cite{Owens,ABFKW,GRV,SMRS}, but pion's 
sea-quark and gluon distributions are only poorly 
constrained. Even less is known about the kaon's partonic structure. The scarce 
experimental information is from the measurement of the $K^-$-induced Drell-Yan 
process~\cite{WA39,NA3}. Based on a total of $\sim 700$ Drell-Yan events, 
the NA3 
Collaboration~\cite{NA3} found the valence $\bar u$  quark 
distribution of $K^-$ to be softer than that of $\pi^-$.
This softer $\bar u$ valence quark distribution in $K^-$ is 
attributed to the presence of the heavier $s$ valence quark, resulting in
a larger fraction of $K^-$'s momentum being carried by the $s$ quark
than the lighter $\bar u$ quark. This indication 
of the flavor dependence 
of the kaon valence quark distributions has been compared with theoretical 
calculations~\cite{Roberts14,Roberts16,Thomas16}. Further experimental 
constraints on the flavor
dependence of the kaon valence quark and gluon distributions are 
clearly of interest.

In this paper, we examine the feasibility to extract information on kaon's
valence quark and gluon distributions using $J/\psi$ production data 
with kaon beams.  
We show that the existing
NA3 data on the $K^- / \pi^-$ ratio for $J/\psi$ production  
provide an independent experimental evidence that 
valence $\bar u$ quark distribution of $K^-$ is softer than that 
of $\pi^-$. The 
$K^+ / \pi^+$ data also suggest that the gluon distribution
in kaon is different from that in pion. 

\begin{figure}[tb]
\includegraphics*[width=\linewidth]{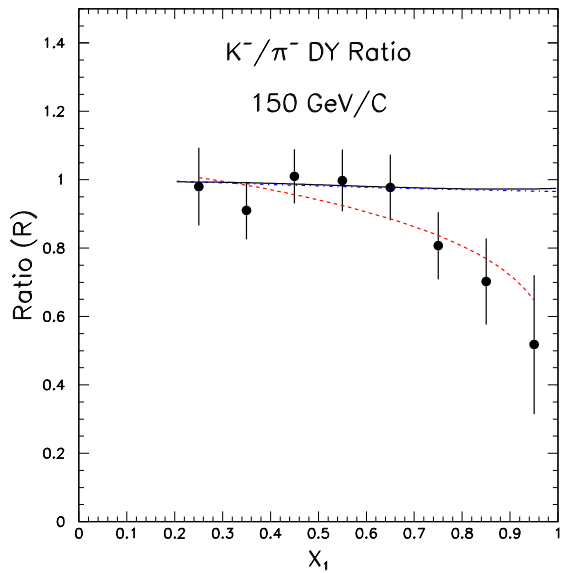}
\caption{Cross section ratio $R$ for $K^-$ and $\pi^-$-induced
Drell-Yan reaction on a platinum target versus $x_1$. 
The data are from NA3~\cite{NA3}
covering the dimuon mass region $4.1 \le M \le 8.5$ GeV utilizing
150 GeV/c meson beams. The solid (dot-dashed) curve corresponds to LO
(NLO) calculation using the SMRS meson 
PDFs for both $K^-$ and $\pi^-$ and the CTEQ5L nucleon PDF. The 
dashed curve corresponds to
calculation with the valence $\bar u$ distribution in $K^-$ multiplied by
a factor of $(1-x_1)^{0.203}$ and the valence $s$ distribution 
divided by the same factor.
The normalizations of the $\bar u$ and
$s$ valence quark distributions are modified to satisfy the valence-quark
number sum rule.}
\label{fig1}
\end{figure}

We first briefly review the experimental evidence for a flavor-dependent
valence quark distribution in kaon. The NA3 Collaboration reported a
simultaneous measurement of $K^- +$ Pt $\to \mu^+ \mu^- + X$ and
$\pi^- +$ Pt $\to \mu^+ \mu^- + X$ using a 150 GeV/c beam on a platinum
target~\cite{NA3}. 
The negatively
charged beam contained $\pi^-, K^-$ and $\bar p$ particles. The particle
type was identified by two Cherenkov counters placed in the beam. 
To select the Drell-Yan events, the masses
of the dimuon events satisfied $4.1 \le M \le 8.5$ GeV. 
Approximately 700 dimuon events produced by $K^-$ and 21,200 events produced
by $\pi^-$ were selected. Figure~\ref{fig1} shows the ratio $R$ for
the dimuon yields measured for $K^-$ and $\pi^-$ beams as a function of
$x_1$, which is the fraction of the beam momentum carried by
the interacting parton. 
The ratio $R$ at large $x_1$ ($x_1 > 0.65$) clearly falls
below unity. Since the Drell-Yan cross sections with $\pi^-$ and
$K^-$ beams at large $x_1$ are dominated by the term containing 
the product of $\bar u_M(x_1)$ in the meson $(M)$
and $u_A(x_2)$ in the nucleus $(A)$, the fall-off in $R$ at large $x_1$ was 
interpreted as an evidence that $\bar u_K(x)$
is softer than $\bar u_\pi (x)$~\cite{NA3}.

To obtain a more quantitative determination on how $\bar u_K(x)$ is modified
with respect to $\bar u_\pi(x)$, we have calculated $R$ using 
the leading-order (LO) 
and next-to-leading-order (NLO) Drell-Yan cross section expression. The solid 
curve in Fig. 1 is a LO calculation
using the SMRS (set 2)~\cite{SMRS}  parton distribution function (PDF)
parametrizations for both $K^-$
and $\pi^-$. The CTEQ5L~\cite{CTEQ5L} parametrization for the
proton PDFs was used for
the platinum target, weighted by the number of protons and neutrons in the
platinum nucleus. The dot-dashed curve represents a calculation
including the NLO
contribution with the DYNNLO code~\cite{DYNNLO}. The nearly identical
results from LO and NLO show that the cross section ratio $R$ is
insensitive to contributions from high-order processes. Thus the
simpler calculation with LO is adequate for $R$.
Figure 1 shows that the data at the large $x_1$ region
are clearly different from
the calculation assuming the same valence-quark distributions for kaon and
pion.

In order to account for the drop of $R$ as $x_1 \to 1$, we assume a
$\bar u^V_K (x)$ distribution softer than $\bar u^V_\pi (x)$ with the
following parametrization:
\begin{eqnarray}
\bar u^V_K (x) = N_u \bar u^V_\pi (x) (1-x)^a.
\label{eq:eq1}
\end{eqnarray}
while the strange valence-quark distribution in $K^-$ is assumed
to be harder than $\bar u^V_\pi (x)$:
\begin{eqnarray}
s^V_K (x) = N_s \bar u^V_\pi (x) / (1-x)^a.
\label{eq:eq2}
\end{eqnarray}
The normalization factors $N_u$ and $N_s$ ensure the 
number sum rules for valence-quark distributions in kaon:
\begin{eqnarray}
\int_0^1 \bar u^V_K(x) dx =1;~~~~\int_0^1 s^V_K(x) dx =1.
\label{eq:eq3}
\end{eqnarray}
Using LO calculation, the best-fit value for
$a$ to describe the NA3 data is found to be $a = 0.203 \pm 0.06$
with the  normalization factors $N_u = 1.061$
and $N_s = 0.937$. 
We assume that $a$ is $Q^2$-independent
in the range of $4.1 < M < 8.5$ GeV.
This assumption is reasonable, since the $Q^2$ dependence in this range is
found to be small for the pion PDFs.
The dashed curve in Fig. 1 corresponds to 
calculation with these best-fit values. The fall-off in $R$ at large
values of $x_1$ is well described by a $\bar u^V_K$ distribution
softer than $\bar u^V_\pi$.

\begin{figure}[tb]
\includegraphics*[width=\linewidth]{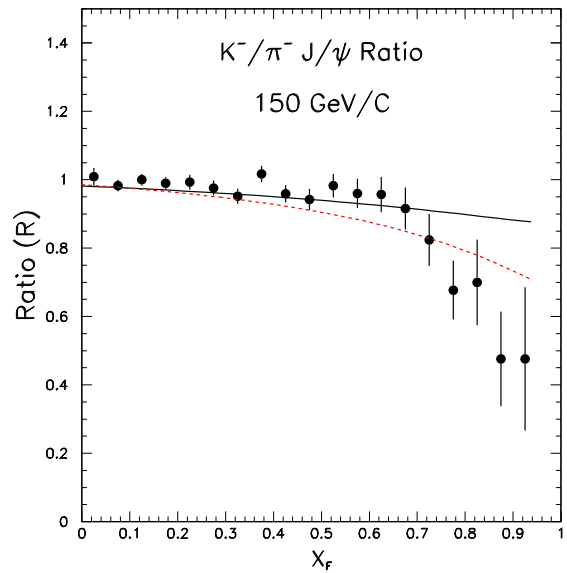}
\caption{Cross section ratio $R$ for $K^-$ and $\pi^-$-induced
$J/\psi$ production on a platinum target versus $x_F$. The data
are from NA3~\cite{NA3_jpsi}. 
The solid curve is a color-evaporation model calculation 
using the SMRS meson 
PDFs for both $K^-$ and $\pi^-$. The dashed curve corresponds to
calculation with the modified valence quark distributions in $K^-$ 
as described in the Fig. 1 caption.}
\label{fig2}
\end{figure}


In addition to the $K^- / \pi^-$ Drell-Yan cross section ratio data
shown in Fig. 1, the NA3 Collaboration also reported the measurement
of $K^- / \pi^-$ ratio for $J/\psi$ production at 150 GeV/c 
on a platinum target~\cite{NA3_jpsi}.
Moreover, the $K^+ / \pi^+$ ratios for $J/\psi$ production have also been
measured at 200 GeV/c~\cite{NA3_jpsi}.
These data have the potential of providing new information on the
parton distributions in kaon. We first investigate the impact of the
$K^- / \pi^-$ ratio for $J/\psi$ production at 150 GeV/c, shown in
Fig. 2. The NA3 data cover a very broad range of the kinematic
variable $x_F$, which is the fraction of the available momentum in
the center-of-mass frame carried by $J/\psi$. While the 
$K^- / \pi^-$ ratio is relatively flat for the region $0 < x_F < 0.6$,
it starts to drop noticeably when $x_F$ further increases.
A comparison between Fig. 1 and Fig. 2 shows a striking similarity.
Since the fall-off at large $x_1$ in the Drell-Yan ratio was described
as a soft $\bar u^V_K$ distribution,
it is possible that the pronounced drop of the $K^- / \pi^-$ ratio
at large $x_F$ in Fig. 2 has a similar origin.

In contrast to the electromagnetic Drell-Yan process, hadronic $J/\psi$
production involves strong interaction for the underlying mechanism.
The two hard processes responsible for producing a pair of heavy
quarks ($c \bar c$ and $b \bar b$) are the $q \bar q$ annihilation and the
$g g$ fusion. The cross section for producing a pair of heavy quarks
($Q \bar Q$) in hadronic interaction is

\begin{eqnarray}
\frac {d\sigma}{dx_F d\tau} = \frac {2 \tau} {(x_F^2+4\tau^2)^{1/2}}
H_{BT}(x_1,x_2;m^2),
\label{eq:eq4}
\end{eqnarray}
where $x_1,x_2$ are the momentum fraction carried by the beam
($B$) and target ($T$) partons, and $x_F = x_1 - x_2$. The mass of 
the $Q \bar Q$ pair is $m$, and $\tau^2 = m^2/s$ where $s$ is the
center-of-mass energy squared. $H_{BT}$ is the convolution of the hard-process
cross sections and the parton distributions of the projectile and 
target hadrons

\begin{eqnarray}
H_{BT}(x_1,x_2;m^2) = G_B(x_1)G_T(x_2)\sigma(gg \to Q \bar Q; m^2) \nonumber \\
+ \sum_{i=u,d,s} [{q^i_B(x_1) \bar q^i_T(x_2)+\bar q^i_B(x_1)q^i_T(x_2)}]
\nonumber \\
\times \sigma(q \bar q \to Q \bar Q;m^2),~~~~~~~~~~~~~~
\label{eq:eq5}
\end{eqnarray}
where $G(x),q(x),$ and $\bar q(x)$ refers to the gluon, quark, and antiquark
distribution functions, respectively. The expressions for the 
cross sections of the QCD subprocesses $\sigma (gg \to Q \bar Q)$ 
and $\sigma (q \bar q \to Q \bar Q)$ can be found in Ref.~\cite{combridge}. 

To go from the production of $c \bar c$ pair to the production of the
$c \bar c$ bound state $J/\psi$, we use the color-evaporation 
model~\cite{cem}. In this model the $c \bar c$ bound state is obtained
by integrating the free $c \bar c$ cross section from the $ c \bar c$ 
threshold, $\tau_1 = 2m_c/\sqrt s$, to the open-charm threshold,
$\tau_2 = 2 m_D/\sqrt s$. The differential cross section for $J/\psi$
production is then given as

\begin{eqnarray}
\frac{d\sigma}{dx_F}=F\int_{\tau_1}^{\tau_2} 2 \tau 
d\tau H_{BT}(x_1,x_2;m^2)/(x_F^2+4\tau^2)^{1/2},
\label{eq:eq6}
\end{eqnarray}
where the factor $F$ accounts for the fraction of the $c \bar c$ 
pairs forming $J/\psi$ either directly or indirectly via the
decay of more massive charmonium states. Despite its simplicity, the
color-evaporation model is capable of describing many salient features of
hadronic $J/\psi$ production~\cite{Barger,Falciano,Vogt,Peng95,Gavai95}. 
For example,
the shape of $\frac{d\sigma}{dx_F}$ as well as their beam-energy and 
beam-type dependencies, which are sensitive to the relative
contributions of $q \bar q$ annihilation and $g g$ fusion, are very well
reproduced. The success of the color-evaporation model suggests that 
this model gives an adequate prediction for the relative contribution 
of these two processes. 

The solid curve in Fig. 2 represents the calculation of the 
$K^-/\pi^-$ ratio for $J/\psi$ production on a platinum target utilizing
the color-evaporation model. The same valence-quark, sea-quark, and gluon
distributions (SMRS PDFs) were assumed for $\pi^-$ and $K^-$, and the 
nucleon PDFs are taken from CTEQ5L. Identical values of $F$ were assumed 
for the $\pi^-$ and $K^-$ cross sections in Eq. (6). Hence the result
is independent of the value of $F$. Figure 2 shows that the color-evaporation
model gives good description of the NA3 data for the region $0 < x_F < 0.65$,
indicating the adequacy of this model. However, at the largest values
of $x_F$ ($0.65 < x_F < 1.0$), the calculation is significantly above 
the data.

\begin{figure}[tb]
\includegraphics*[width=\linewidth]{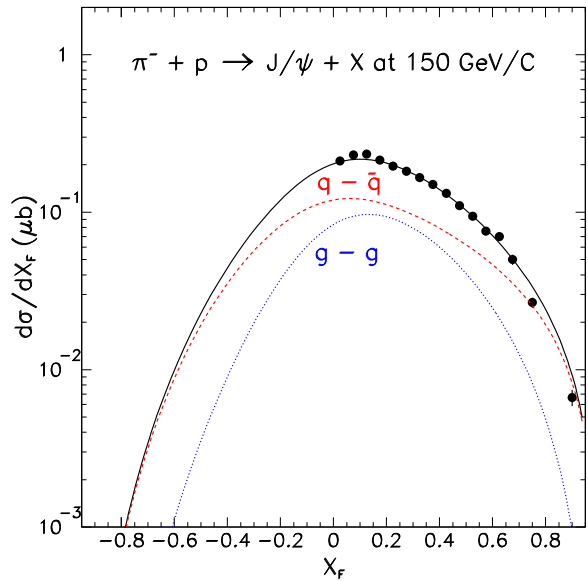}
\caption{Measured cross section and calculations 
for $\pi^-$-induced
$J/\psi$ production on a hydrogen target versus $x_F$.
The data are from NA3~\cite{NA3_jpsi,NA3_thesis} and the calculation 
uses the color-evaporation model.
The SMRS meson PDFs for $\pi^-$,
together with the CTEQ5L nucleon PDF are used in the
calculation. The factor $F$ in Eq. (6) is 0.325.
The contributions
fron the $q - \bar q$ annihilation and the $g - g$ fusion are
shown as the dashed and dotted curve, respectively.}
\label{fig3}
\end{figure}

To shed some light on the origin of the discrepancy between the data
and the calculation shown in Fig. 2, we show in Fig. 3 the comparison
between the $\pi^- +$ p $J/\psi$ production data from 
NA3~\cite{NA3_jpsi,NA3_thesis} with the color-evaporation
model calculation, using the same PDF inputs as in Fig. 2. The factor $F$ 
for the solid curve is $F=0.325$, resulting in a very good agreement
with the data. This suggests that the discrepancy 
observed in Fig. 2 originates from the calculation for $K^- +$ Pt,
rather than $\pi^-$ + Pt. Additional information is provided
in Fig. 3 where the contributions from the $q \bar q$ annihilation and the
$g g$ fusion are shown as the dashed and dotted curve, respectively. 
One notes that the $q \bar q$ annihilation process becomes increasingly 
important at the large
$x_F$ region relative to the $g g$ fusion process. This
implies that the $K^-/\pi^-$ $J/\psi$ ratio at large $x_F$ should be
sensitive to the valence quark distribution in $K^-$, and
the drop in $R$ could reflect a softer $\bar u$ valence-quark
distribution in $K^-$ than in $\pi^-$.
We have carried out color-evaporation model calculation, shown as 
the dashed curve in
Fig. 2, using the modified kaon valence-quark
PDFs as in Eqs. (1) and (2) with $a=0.203$. The modified kaon PDFs
give a better description of the data. However, the data at the
largest $x_F$ region remain slightly below the calculation.
It would be very helpful to obtain new high statistics dimuon data at 
the forward
$x_F$ region with kaon beam. A future global analysis of both the Drell-Yan
and $J/\psi$ production data would allow an extraction of the quark and gluon
distributions in kaon.
 
\begin{figure}[tb]
\includegraphics*[width=\linewidth]{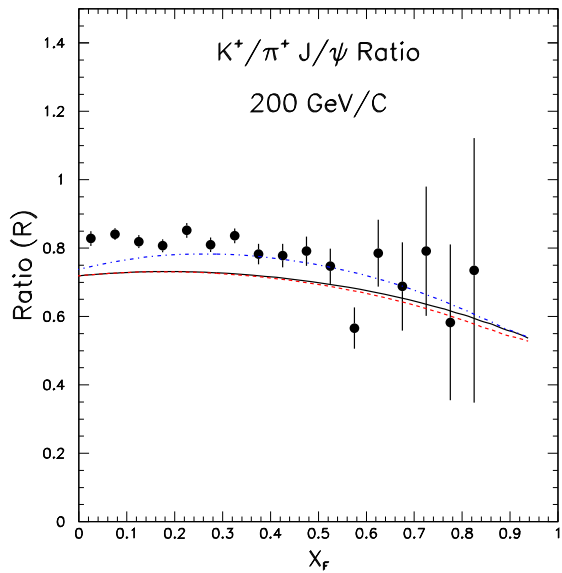}
\caption{Cross section ratio $R$ for $K^+$ and $\pi^+$-induced
$J/\psi$ production on platinum target versus $x_F$.
The solid curve is a calculation with the color-evaporation model
using the SMRS meson
PDFs for both $K^+$ and $\pi^+$,
together with the CTEQ5L nucleon PDF. The dashed curve corresponds to
calculation with the valence quark distribution in $K^+$ modified as described
in the Fig. 1 caption. The dot-dashed curve corresponds to a calculation where
the gluon snd sea-quark distributions in kaon are modified, as 
discussed in the text.}
\label{fig4}
\end{figure}

The NA3 Collaboration has also measured the $K^+ / \pi^+$ ratios for 
$J/\psi$ production at 200 GeV/c 
on a platinum target as shown in Fig. 4~\cite{NA3_jpsi}.
A striking difference between the $K^+ / \pi^+$ and the 
$K^- / \pi^-$ ratios is observed. While there is a pronounced drop
of the $K^- / \pi^-$ ratios at forward $x_F$, no such drop is present
for $K^+ / \pi^+$. The solid curve in Fig. 4 is the color-evaporation
model calculation
using the same PDFs for $\pi^+$ and $K^+$. As expected, the calculation
shows only a mild drop at the large $x_F$ region, similar to the data.
The dashed curve in Fig. 4
is obtained with a modified $K^+$ PDFs following Eqs. (1) and (2).
It is worth noting that the modified $K^+$ PDFs give a very similar
result. The insensitivity of 
$K^+$ $J/\psi$ production to the
valence quark distribution in $K^+$ reflects the fact that 
the $u$ and $\bar s$ valence quarks in $K^+$ must 
annihilate, respectively, 
$\bar u$ and $s$ sea quarks in the nuclei. Unlike the $K^- +$ Pt reaction,
where the $\bar u$ valence quark can annihilate a $u$ valence quark
in nucleus, the $q \bar q$ annihilation in the $K^+ +$ Pt reaction
must involve at least one sea quark from the beam and the target nucleus.  
Hence, the $q \bar q$
annihilation contribution is much suppressed when compared with
the $g g$ fusion contribution. This leads to the striking difference
between the $x_F$ dependence of the $K^+/\pi^+$ and $K^-/\pi^-$
$J/\psi$ production ratios.

Figure 4 shows that the $K^+/\pi^+$ $J/\psi$ data are
higher than the calculation by roughly 10-15\%. This reflects the
greater importance of $g g$ fusion due to the suppression of $q \bar q$ 
annihilation in the $K^+$ + Pt $J/\psi$ production. 
Since the gluon distribution
is assumed to be the same in $\pi^+$ and $K^+$, the difference between the
data and the calculation may suggest a difference in the gluon distribution
between $\pi^+$ and $K^+$. Note that a recent paper indeed suggests different
gluon distributions for pion and kaon~\cite{Roberts16}. To examine the
sensitivity of the $K^+/\pi^+$ $J/\psi$ cross section ratio to the gluon
distributions, the dot-dashed curve in Fig. 4 corresponds to a calculation
in which the gluon distribution in kaon at the $J/\psi$ mass scale is increased
by 12\% so that 45\% of kaon's momentum is carried by gluons, rather than
the nominal value of 40\% in SMRS (set 2). To ensure the momentum sum
rule for kaon, the sea-quark distribution is decreased accordingly 
so that 10\% of 
kaon's momentum
is carried by the sea-quarks, rather than the nominal value of 15\%.
The modified gluon and sea-quark distributions in kaon maintain the same
functional form in $x$, but with different normalizations. Figure 4 shows that
the $K^+/\pi^+$ $J/\psi$ data favor a larger gluon content in kaon
relative to pion.  

In summary, we have investigated the feasibility to extract information
of the kaon valence quark distribution via the $J/\psi$ production
data. We show that the existing data from NA3 provide an independent
evidence that the $\bar u$ valence quark distribution of $K^-$ has a
softer $x$ distribution than that of $\pi^-$. This is in qualitative
agreement with the result obtained from the analysis on the 
$K^- / \pi^-$ Drell-Yan cross section ratio data. Since $J/\psi$ production
proceeds via strong interaction with a cross section much larger than 
that of the electromagnetic Drell-Yan process, it is a promising independent
experiment tool to extract the partonic distributions in the kaon. 
As the $J/\psi$ production also involves the $g g$ fusion contribution,
it is conceivable that one could also extract unique information on the
gluon distribution in kaon, which is completely unknown
at this moment. The prospect for pursuing Drell-Yan and $J/\psi$
production with an intense RF-separated kaon beam is under active
consideration~\cite{ect}.

This work was supported in part by the U.S. National Science Foundation
and the Ministry of Science and
Technology of Taiwan.


\begin{thebibliography}{99}
\bibitem{Roberts_Holt} R. Holt and C.D. Roberts, Rev. Mod. Phys. {\bf 82},
2991 (2010).
\bibitem{DY} S.D. Drell and T.M. Yan, Phys. Rev. Lett. {\bf 25}, 316
   (1970); Ann. Phys. (NY) {\bf 66}, 578 (1971).
\bibitem{Kenyon} I.R. Kenyon, Rep. Prog. Phys. {\bf 54}, 1261 (1982).
\bibitem{WA39} M. Corden {\em et al.}, Phys.
Lett. B {\bf 96}, 417 (1980).
\bibitem{NA3} J. Badier {\em et al.}, Phys.
Lett. B {\bf 93}, 354 (1980).
\bibitem{Chang13} W.C. Chang and D. Dutta, Int. J. Mod. Phys. E {\bf 22},
1330020 (2013).
\bibitem{WA70} M. Bonesini {\em et al.}, Z. Phys. C
{\bf 37}, 535 (1988); {\bf 38}, 371 (1988).
\bibitem{Owens} J.F. Owens, Phys. Rev. D {\bf 30}, 943 (1984).
\bibitem{ABFKW} P. Aurenche et al., Phys. Lett. B {\bf 233}, 517 (1989).
\bibitem{GRV} M. Gl\"{u}ck, E. Reya, and A. Vogt, Z. Phys. C 
{\bf 53}, 651 (1992)
\bibitem{SMRS} P.J. Sutton, A.D. Martin, R.G. Roberts, 
and W.J. Stirling, Phys. Rev. D {\bf 45}, 2349 (1992).
\bibitem{Roberts14} L. Chang {\em et al.}, Phys. Lett. B {\bf 737},
23 (2014).
\bibitem{Roberts16} C. Chen {\em et al.}, Phys. Rev. D {\bf 93}, 074021
(2016).
\bibitem{Thomas16} P. Hutauruk {\em et al.}, Phys. Rev. C {\bf 94}, 035201
(2016).
\bibitem{NA3_jpsi} J. Badier {\em et al.}, Z. Phys. C
{\bf 20}, 101 (1983).
\bibitem{CTEQ5L} H.L. Lai {\em et al.}, Eur. Phys. J. C {\bf 12}, 375 (2000).
\bibitem{DYNNLO} S. Catani {\em et al.}, Phys. Rev. Lett. {\bf 103}, 
082001 (2009).
\bibitem{combridge} B.L. Combridge, Nucl. Phys. B {\bf 151}, 429 (1979).
\bibitem{cem} H. Fritzsch, Phys. Lett. B {\bf 67}, 217 (1977).
\bibitem{Barger}  V. Barger, W.Y. Keung, and R.J.N. Phillips, Z. Phys. C 
{\bf 6}, 169 (1980).
\bibitem{Falciano}  S. Falciano {\em et al.}, Phys. Lett. B {\bf 158}, 
92 (1985).
\bibitem{Vogt} R. Vogt, S.J. Brodsky, and P. Hoyer, Nucl. Phys. B 
{\bf 360}, 67 (1991).
\bibitem{Peng95} J.C. Peng, D.M. Jansen, and Y.C. Chen, Phys. Lett. B
{\bf 344}, 1 (1995).
\bibitem{Gavai95} R. Gavai {\em et al.}, Int. J. Mod. Phys. 
A {\bf 20}, 3043 (1995).
\bibitem{NA3_thesis} Ph. Charpentier, Thesis, Universit\`{e} Paris-Sud,
unpublished (1983).
\bibitem{ect} Workshop on ``Dilepton Production with Meson and Antiproton
beams", ECT$^*$ Workshop, Nov. 2017, http://www.ectstart.eu/node/2232. 
\end{thebibliography}
\end{document}